\documentclass[aps,prl,twocolumn,showpacs,superscriptaddress,groupedaddress]{revtex4}

\usepackage{graphicx}  
\usepackage{epsfig}
\usepackage{dcolumn}   
\usepackage{bm}        
\usepackage{amssymb}   
\usepackage{amsmath}
\begin{document}


\title{The `Higgs' amplitude mode in weak ferromagnetic metals}

\author{
Yi Zhang$^1$, Paulo F. Farinas$^2$, and Kevin S. Bedell$^1$\\
\textit{$^1$Department of Physics, Boston College, Chestnut Hill,
Massachusetts 02467, USA}\\
\textit{$^2$Departamento de F\' \i sica, Universidade Federal de S\~ ao
Carlos,
13565-905, S\~ ao Carlos, SP, Brazil}
}

\date{\today}

\begin{abstract}
Using Ferromagnetic Fermi liquid theory, Bedell and Blagoev derived
the collective low-energy excitations of a weak ferromagnet.  They
obtained the well-known magnon (Nambu-Goldstone) mode and found a new
gapped mode that was never studied in weak ferromagnetic metals.  In
this article we have identified this mode as the Higgs boson (amplitude
mode) of a ferromagnetic metal.  This is identified as the Higgs since
it can be show that it corresponds to a fluctuation of the amplitude
of the order parameter.   We use this model to describe the
itinerant-electron ferromagnetic material MnSi.  By fitting the model
with the existing experimental results, we calculate the dynamical
structure function and see well-defined peaks contributed from the
magnon and the Higgs.  Our estimates of the relative intensity of
the Higgs amplitude mode suggest that it can be seen in neutron
scattering experiments on MnSi.

\end{abstract}
\maketitle

The emergence of low-energy excitations in systems with spontaneously
broken symmetry plays a very important role in our fundamental
understanding of nature. There are two types of fundamental
excitations (particles) that may be present in the field theory
descriptions of these spontaneously broken symmetries: massless
Nambu-Goldstone bosons (magnons, or phase modes)
and massive Higgs bosons (amplitude
modes). Let the order parameter that
describes a state with a spontaneously broken symmetry be $\phi({\bf
x})=\rho({\bf x})e^{i\theta({\bf x})}$. Then, in
the ground state we have $<\phi({\bf x})>=\rho_0 e^{i\theta_0}$,
where finite values for $\theta_0$ and $\rho_0$ correspond to a
broken symmetry. The two fundamental excitations are the
Nambu-Goldstone boson, which corresponds to fluctuations of the phase
of the order parameter with fixed amplitude, $\rho_0$, and the Higgs
boson, with fluctuations of $\rho({\bf x})$ with the phase fixed at
$\theta_0$.

The understanding of the amplitude mode is particularly important for
the Standard model of elementary particle physics\cite{W}. Recent
experiments from LHC\cite{LHC1,LHC2} which detected Higgs-like
particles has drawn much attention to this subject. This Higgs-like
mode was predicted theoretically and found experimentally in many
condensed matter systems. This includes incommensurate
charge-density-wave (CDW) states\cite{LRA}, superconducting
systems\cite{SK,LV1,LV2,V}, and lattice Boson systems with superfluid
and Mott insulator transition\cite{PAP,AA,PP,HABB,Nature}, which can
be realized by ultracold bosonic atoms on an optical lattice.

In this article we have identified the amplitude mode in another
well-known system with spontaneously broken symmetry, a ferromagnetic
metal. The Nambu-Goldstone magnon mode in ferromagnetic systems was
first predicted by Bloch\cite{B} and Slater\cite{S}, and observed in
iron in a neutron scattering experiment\cite{L}. Abrikosov and
Dzyaloshinkii\cite{AD} first predicted spin waves in ferromagnetic
Fermi liquids using a ferromagnetic Fermi liquid theory (FFLT) for
itinerant ferromagnets. This approach was put on a more microscopic
foundation in the work of Dzyaloshinskii and Kondratenko\cite{DK}.
Moriya and Kawabata also developed a band theory (MK theory) to study
the long wavelength spin fluctuation in itinerant ferromagnetic
systems\cite{MK}. Bedell and Blagoev\cite{BB} generalized the FFLT and
they discovered a new collective mode of the system; they found a mode
with a gap in the excitation spectrum.  We have identified this mode
as the Higgs amplitude mode of a ferromagnetic metal.

We begin by quickly describing the FFLT approach used by Bedell and
Blagoev \cite{BB}, in the small magnetic moment limit, to study the
collective excitations in a ferromagnetic metal.
We then arguee that the massive mode is indeed an amplitude mode.
To estimate the collective modes' frequencies we
use a simple one-band description with a spherical Fermi surface
for the itinerant-electron ferromagnet MnSi.  By fitting
existing experimental results we can pin down some of
the Landau parameters and use them to calculate dynamical structure
function, $S(q,\omega)$.  As we will see there are two
sharp peaks in $S(q,\omega)$ corresponding to the two
collective modes, the well-known Nambu-Goldstone mode (sometimes
referred to as the transverse spin wave) and a gapped Higgs amplitude
mode.  From our estimates of the spectral weight of the amplitude mode
it should be observable in neutron scattering experiments.

We consider a three dimensional weak ferromagnetic material below its
Curie temperature. According to FFLT, in the weak moment limit, the
system can be described by the quasi-particle distribution function
$n_{{\bf p}\alpha\alpha^{'}}(\vec r,t)=n_{\bf p}({\bf
r},t)\delta_{\alpha\alpha^{'}}+{\bf m_p}({\bf
r},t)\cdot\vec\sigma_{\alpha\alpha^{'}}$,
and the quasi-particle energy
function
$\epsilon_{{\bf p}\alpha\alpha^{'}}(\vec r,t)=\epsilon_{\bf p}({\bf
r},t)\delta_{\alpha\alpha^{'}}+{\bf h_p}({\bf
r},t)\cdot\vec\sigma_{\alpha\alpha^{'}}
$\cite{BP,AD},
with ${\bf m_p}$ the quasi-particle magnetization and ${\bf h_p}=-{\bf
B}+2\sum_{\bf p'}f_{\bf p p'}^a {\bf m_p'}$ the effective magnetic
field which includes the external magnetic field $\bf B$ and the
internal magnetic field generated by quasi-particle interactions
$f_{\bf pp^{'}}^a$. Here and throughout the paper we set
$\mu_{mag}=\frac{\gamma\hbar}{2}=1$ and $\hbar=1$. The quasi-particle
interaction can be expanded in Legendre Polynomials, and in this weak
moment limit, it can be treated as spin rotation invariant\cite{BEB},
so that it can be separated into the spin symmetric and spin
anti-symmetric parts
$N(0)f_{\bf
pp^\prime}^{\sigma\sigma^\prime}
=\sum_l(F_l^s+F_l^a\sigma\cdot\sigma^\prime)
P_l(\hat{\bf p}\cdot\hat{\bf p}^\prime).
$
Here, $N(0)$ is the average density of states over the two Fermi
surfaces and in the ferromagnetic phase, $F_0^a<-1$. This state can be
said to be protected by the generalized Pomeranchuck stability
condition\cite{P,YB},
and its ground state is described by
the distribution function
$
{\bf m}_{\bf p}^{0}=-{\bf m}_0\partial n_{{\bf
p}}^{0}/\partial\varepsilon_{{\bf p}}^0,
$
where ${\bf m}_0$ is the equilibrium magnetization divided
by $N(0)$. Fluctuations about equilibrium, $\delta {\bf m_p}$,
are described by
a linearized spin kinetic equation,\cite{BP}
\begin{widetext}
\begin{equation}
\frac{\partial {\delta\bf m_p}({\bf r},t)}{\partial t}+{\bf v_p}\cdot
{\bf\nabla}(\delta{\bf m_p}({\bf r},t)-\frac{\partial n_{\bf
p}^0}{\partial\varepsilon_{\bf p}^0}\delta{\bf h_p}({\bf
r},t))\\=-2({\bf m}_{\bf p}^0({\bf r},t)\times \delta{\bf h}_{\bf
p}({\bf r},t)+\delta{\bf m_p}({\bf r},t)\times{\bf h}_{\bf p}^0({\bf
r},t))+I[{\bf m_p}]
\label{kinetic}
\end{equation}
\end{widetext}
where ${\bf h_p^0}=-{\bf B}+2\sum_{\bf p'}f_{\bf p p'}^a {\bf m_p'}$ and $\delta{\bf h_p}=-\delta{\bf B}+2\sum_{\bf p'}f_{\bf p p'}^a\delta{\bf m_p'}$ are the effective equilibrium field and its fluctuation, respectively. Here, a small magnetic field $\delta{\bf B}$ is set transverse to the equilibrium magnetization.
A direct way to obtain the dispersions is to take
the limit of free oscillations of Eq.(\ref{kinetic}) by setting
${\bf B}=0$ and ${\delta\bf B}=0$, and looking at the
low temperature limit for which the collision integral $I[{\bf m_p}]$
can be ignored. This is the quantum spin
hydrodynamic regime whose details of the
derivation are more explicit in Ref.\onlinecite{BB}.
In addition to the free oscillation limit, it
is also possible to use Eq.(\ref{kinetic})
to calculate the response function $\chi(q,\omega)$
that relates ${\delta\bf B}$ with its induced transverse
response $\delta{\bf m}=\sum_{\bf p}\delta{\bf m_ p}$
($\delta{\bf m}=\chi(q,\omega)\delta{\bf B}$) and from
it the dynamical structure function $S(q,\omega)=-Im[\chi(q,\omega)]/\pi$.
The calculations are done by projecting out the $l$ components of
both the kinetic
equation and the spin density
distribution function $\delta{\bf m_p}=
-(\partial n_{\bf
p}^0/\partial\varepsilon_{\bf p}^0)
\sum_{l}{\bm \nu}_{l}P_l(\hat{\bf
p}\cdot\hat{z})$.
We will return to the full calculation of $S(q,\omega)$
a little later.

We can use the kinetic equation to derive the continuity equation for
the magnetization and the equation of the spin current defined as
${\bf j}_i^{\sigma}({\bf r},t)=\sum_{\bf p}v_{pi}{\bf m_p}({\bf
r},t)(1+F_1^a/3)$. The dispersions found using the free-
oscillation limit are
\begin{eqnarray}
\omega_1^\pm(q)=\frac{c_s^2}{\omega^\pm}q^2,
\label{dis1}
\\
\omega_2^\pm(q)=\omega^\pm-\frac{c_s^2}{\omega^\pm}q^2
\label{dis2}
\end{eqnarray}
where $\omega^\pm=\pm 2m_0\left|F_0^a-F_1^a/3\right|$,
$c_s^2=\left|1+F_0^a\right|(1+F_1^a/3)v_F^2/3$,
and $v_F$ is the Fermi velocity on the average Fermi surface.
The $\pm $ signs correspond simply to the different precessional
directions so that we effectivelly have two modes.

\begin{figure}[ht]
\begin{center}
\includegraphics[scale=0.4]{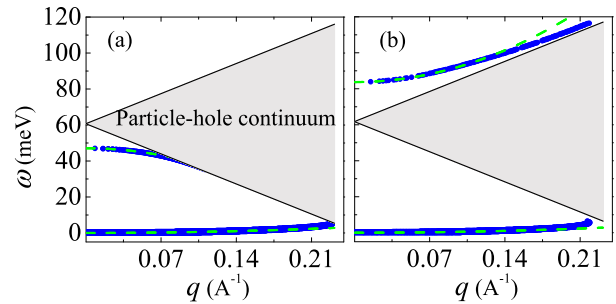}
\end{center}
\caption{(color on line)
Collective modes together with the p-h continuum in the case
(a) $F_0^a=-1.18$, $F_1^a=-0.84$ and (b) $F_0^a=-1.16$, $F_1^a=1.32$. 
The green-dashed lines represent the dispersion calculated using the
hydrodynamic approach, Eqs.(\ref{dis1}) and (\ref{dis2}) while
the blue-solid lines
are the dispersions taken from the poles of the response
function. Plots obtained by using parameters fitted to
early experiments (see text), including the following:
$N(0)=6.4\times10^{29}\;{\rm eV^{-1} m^{-3}}$, $k_F=1.23\;{\rm \AA}^{-1}$,
$m^*=39.3m_e$, $E_F=0.147\;{\rm eV}$, and $v_F=3.63\times10^4\;{\rm m
s^{-1}}$.}\label{fig:figure1}
\end{figure}

Before we go to the full calculation of $S(q,\omega)$,
we can extract the basic physics
from these results. The `Standard Model' of a ferromagnetic metal is
often referred to as the Stoner model and it can best be characterized
by its elementary excitations. To begin with there are the spin 1/2
particle-like excitations (sometimes referred to as quasi-particles),
consisting of up spins ($\sigma=\uparrow$ ), majority spins, and down
spins ($\sigma=\downarrow$ ), minority spins. In the presence of
spontaneous long range ferromagnetic order there are also gapless
transverse spin waves (the Nambu-Goldstone mode). This mode has a
total spin of $\pm 1$ associated with the two precessional
directions of the equilibrium magnetization.
For simplicity, in what follows we will
consider only the spin $+1$ excitations. If we think in terms of the
order parameter for the ferromagnetic state this would be a
fluctuation of the phase (rotation about the z axis in space) with the
magnitude of the order parameter fixed at its equilibrium value,
$m_0$. In addition to these spin waves,
there are other spin $+1$ excitations in the Stoner model that
correspond to fluctuations in the magnitude of the order parameter at
a fixed phase. At $q=0$ these excitations have a gap in their
spectrum usually referred to as the Stoner gap.
For $q\ne 0$, these spin $+1$ excitations have a range of
frequencies (shaded region shown in Fig.\ref{fig:figure1})
and these are the incoherent
particle-hole excitations, $\omega_{p-h}^\pm=\mp 2m_0 F_0^a+{\bf
q}\cdot{\bf v_p}$. These excitations are not collective, thus, there
is no Higgs amplitude mode in the Stoner model: When the
momentum-transfer, $q$, is large enough the Goldstone mode decays
into these incoherent particle-hole excitations (Landau damping),
while the gapped exitations are Landau damped for all values
of $q$.

The FFLT description of Bedell and Blagoev[20] for small momentum
transfers is qualitatively the same as the Stoner model if we set all
Landau parameters, $F_l^a=0$, for all $l>0$, and keep only $F_0^a$.
If we keep only the $l=0$ and $l=1$ moments of the spin density
distribution function ($\nu_0$ and $\nu_1$), we get the two modes,
Eqs.(\ref{dis1}) and (\ref{dis2}).
The first mode is just the Nambu-Goldstone mode and the
second mode has a gap in its spectrum, where at $q=0$, it is just the
Stoner gap, $\omega_2^+=2\left|F_0^a\right| m_0$. This excitation
causes a change in the magnitude (amplitude) of the order parameter
since it is a spin flip process.  In the spin flip process we take a
down spin and flip it to an up spin causing an amplitude fluctuation
since we decreased the number of down spins while increasing the
number of up spins during this fluctuation. These fluctuations of the
amplitude of the order parameter could have been the Higgs amplitude
mode, however, in the Stoner model this mode sits in the particle-hole
continuum and it is Landau damped; it is not a collective mode.

The Fermi liquid description of the collective modes of a
Ferromagnetic metal\cite{BB} goes beyond the Stoner model described
above in a simple but most important way.  As can be seen from
Eqs.(\ref{dis1}) and (\ref{dis2}) there are two modes,
the first one is the Goldstone mode.  The
second mode has a gap in its spectrum given by
$\omega_2^+=2m_0\left|F_0^a-F_1^a/3\right|$. The introduction
of the higher order Fermi liquid parameter $F_1^a$ is
responsible for the propagation of the mode. This parameter couples
the momentum of the quasi-particle to its spin and it is responsible
for pushing the mode out of the particle-hole continuum. Most of the
spectral weight in this propagating mode comes from the incoherent
particle-hole continuum. As we noted earlier the particle-hole
continuum is made up from incoherent spin flip (spin $+1$) excitations
and they correspond to an amplitude fluctuation. This mode can sit
above the Stoner gap for positive $F_1^a$ and below the Stoner gap for
negative $F_1^a$, with the lower bound, $F_1^a>-3$\cite{YB}.
It propagates and is built out of fluctuations that change the
amplitude of the order parameter, thus it is the ferromagnetic metal
example of the Higgs amplitude mode. Also since it couples to the
spin fluctuations it can be seen in the transverse spin fluctuation
response function that could be measured in a neutron scattering
experiment, as we see below.

We calculate the dynamical spin-spin response function based on the
kinetic equation, Eq.(\ref{kinetic}). Here we go
beyond the hydrodynamic approach by
keeping ${\bm \nu}_l$ for arbitrary values of $l$ and
the Landau parameters $F_l^a$ up to $l=1$. The
response function obtained is given by
\begin{equation}
\frac{\chi(q,\omega)}{N(0)}
=\frac{-\chi_0^++\frac{2m_0F_1^a}{q
v_F(1+F_1^a/3)}\chi_1^+}{1-F_0^a\chi_0^+-\frac{\omega}{q
v_F}\frac{F_1^a}{1+F_1^a/3}\chi_1^+}
\label{chi}
\end{equation}
where
\begin{eqnarray}
\begin{split}
&\chi_0^+(q,\omega)= -1\\&\ +\frac{1}{2q
v_F}[\omega+2m_0(1+F_0^a)]\ln
{\left(\frac{\omega+2m_0F_0^a+q
v_F}{\omega+2m_0F_0^a-q v_F}\right)},
\end{split}
\\
\begin{split}
&\chi_1^+(q,\omega)=-\frac{1}{q
v_F}[\omega+2m_0(1+F_0^a)]\\&\times[1-\frac{1}{2q
v_F}(\omega+2m_0F_0^a)\ln
\left(\frac{\omega+2m_0F_0^a+q
v_F}{\omega+2m_0F_0^a-q v_F}\right)].
\end{split}
\end{eqnarray}

\begin{figure}[h]
\begin{center}
\includegraphics[scale=0.35]{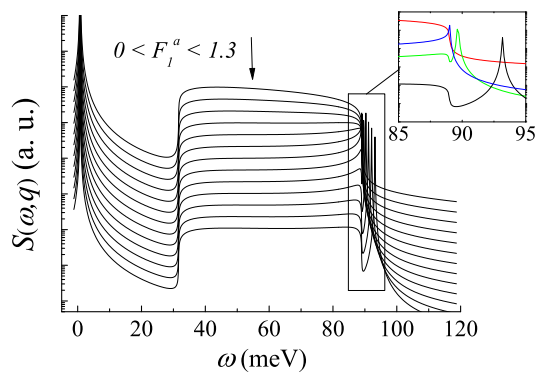}
\end{center}
\caption{(color on line)
Dynamic structure function for different values of
of $F_1^a$. As $F_1^a$ is switched on, the Higgs
builds out of the particle-hole continuum (the
wider peak in the middle) while the Goldstone
mode remains approximately unchanged.}\label{fig0}
\end{figure}

From the response function, we obtain the dynamic structure
function $S(q,\omega)=-Im[\chi(q,\omega)]/\pi$,
whose poles will give,
for small $q$s,
$\omega_1^+(q)$ and
$\bar{\omega}_2^+(q)=\omega^+-(\frac{c_s^2}{\omega^+}-\alpha)q^2$ with
$\alpha=2v_F^2(1+3/F_1^a)/15m_0$, corresponding to the two collective modes derived in hydrodynamic approach respectively. In [20] it was
noted that ${\bm \nu}_2$ is of the same order of ${\bm \nu}_1$ while
${\bm \nu}_l\ll{\bm \nu}_2$ for all $l\geq 3$. In our calculation
we look for
the pole of the response function, which includes all ${\bm \nu}_l$s.
A comparison to the modes obtained in Ref.\onlinecite{BB}
with the hydrodynamic approach is
illustrated in Fig.(\ref{fig:figure1}).
It shows, as
expected, agreement of the two results over a
considerable range of small values for $q$. However,
the full calculation of $S(q,\omega)$ captures
additional features, as, e.g., seen in Fig.(\ref{fig0})
where the value of $F_1^a$ is varied to illustrate
the building of the Higgs mode out of the particle-hole continuum.
As we noted earlier, all of the spectral strength for the Higgs mode
comes from the continuum, which makes the observation
of the continuum itself difficult, as it
is known from neutron scattering experiments.

The itinerant ferromagnetic material MnSi is a very good
candidate for the experimental search of the Higgs mode.
MnSi is metallic
and it behaves ferromagnetically in a magnetic
field\cite{WWSW,SA,WWS,LLS}. Neutron scattering leads to the magnon
dispersion\cite{IST}, $\omega\;({\rm meV})=0.13+52q^2\;({\rm \AA^{-2}})$
and specific
heat measurement gives\cite{CV1,CV2} $(C/T)_0=85*10^-4\;{\rm cal/(K^2
mole})$. From the band structure calculation\cite{JP}, we know the
density of the quasi-particles. We fit the experimental data to
a single band description with the
quadratic dispersion $E=\hbar^2k^2/2m^*$, which keeps the
volume of the Fermi surface unchanged. We obtain
all the parameters we need and a relationship between two of the
Landau parameters, $F_1^a=-(375+321F_0^a)/(143+125F_0^a)$. Since
the system is weakly ferromagnetic, $F_0^a$ should be close to and
smaller than $-1$, and $F_1^a$ depends on $F_0^a$ very sensitively in
this region. We don't have extra experimental results to pin down
the sign of $F_1^a$, so we take the values of ($F_0^a$, $F_1^a$) to be
(-1.16, 1.32) and (-1.18,-0.84) as typical examples to show the two
collective modes in these two cases.
\begin{figure}[ht]
\begin{center}
\includegraphics[scale=0.4]{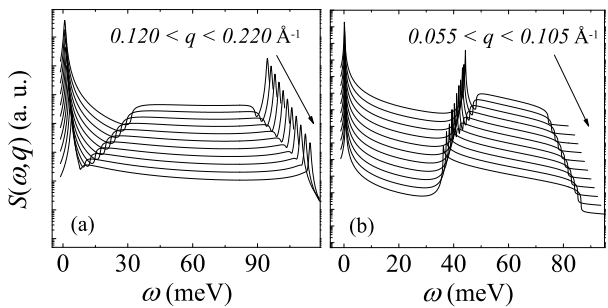}
\end{center}
\caption{Dynamical structure function showing both the collective
modes and the p-h continuum in the cases of $F_0^a=-1.16,$
$F_1^a=1.32$ (a) and $F_0^a=-1.18$, $F_1^a=-0.84$ (b).
}\label{fig:figure2}
\end{figure}

Fig.(\ref{fig:figure2})
shows the dynamic structure function in the two typical cases
for different momentum transfer. We can clearly see the two
sharp peaks contributed by the two collective modes and the
wider one coming from the p-h continuum.

In summary, following from the earlier work of Bedell and
Blagoev\cite{BB}, we used the ferromagnetic Fermi liquid theory to
study the collective modes in a weak ferromagnetic metal. In addition
to the well-known magnon (the phase mode), a gapped mode was also
found\cite{BB}. We have shown here that this gapped mode corresponds
to the Higgs amplitude mode. This mode sits close to the Stoner gap
and is propagating at small $\bf q$ and becomes Landau damped at
larger $\bf q$. We believe that this is the first time that the Higgs
amplitude mode has been predicted in a weak ferromagnetic metal. We
believe the itinerant weak ferromagnet MnSi is a good candidate to
search for this mode and that it should be visible in inelastic
neutron scattering experiments.

We would like to thank Stephen Wilson and Krastan Blagoev for valuable
discussion and advice.

\end{document}